\documentclass[twocolumn,amsmath,amssymb]{revtex4}

\usepackage{graphicx}
\usepackage{dcolumn}
\usepackage{bm}
\usepackage{epstopdf}
\usepackage{multirow}
\usepackage{color}
\usepackage{ifpdf}
\usepackage{amsmath}
\usepackage{amssymb}
\usepackage{epsfig}
\usepackage{graphicx}

\begin{document}

\title{
The pace of evolution across fitness valleys
}

\affiliation{Emmy--Noether Group for Evolutionary Dynamics, Department of Evolutionary Ecology,
Max-Planck-Institute for Evolutionary Biology, August-Thienemann-Str. 2, D--24306 Pl\"on, Germany\\
}

\author{Chaitanya S.~Gokhale $^{\rm 1}$, Yoh Iwasa $^{\rm 2}$, Martin A.~Nowak $^{\rm 3}$, and Arne Traulsen $^{\rm 1}$ 
}
\affiliation{{\small{{\bf{1}} Max-Planck-Institute for Evolutionary Biology, August-Thienemann-Str. 2, 24306 Pl\"{o}n, Germany, {\bf{2}} Department of Biology, Faculty of Sciences, Kyushu University, Fukuoka 812-8581, Japan, {\bf{3}} Program for Evolutionary Dynamics, Department of Mathematics, Department of Organismic and Evolutionary Biology, Harvard University, Cambridge MA 02138, USA}}}  

\begin{abstract}
{
How fast does a population evolve from one fitness peak to another?
We study the dynamics of evolving, asexually reproducing populations in which a certain number of mutations jointly confer a fitness advantage. 
We consider the time until a population has evolved from one fitness peak to another one with a higher fitness.
The order of mutations can either be fixed or random. 
If the order of mutations is fixed, then the population follows a metaphorical ridge, a single path. 
If the order of mutations is arbitrary, then there are many ways to evolve to the higher fitness state. 
We address the time required for fixation in such scenarios and study how it is affected by the order of mutations, the population size, the fitness values and the mutation rate. 
}
\end{abstract}

\maketitle

\section{Introduction}

Evolutionary dynamics is based on natural selection, mutation and genetic drift
\cite{nowak:2006bo}.
It can be illustrated as the dynamics of a population in an abstract, typically
high-dimensional fitness landscape.  
Since individuals with higher fitness produce more offspring, the average density of
individuals is highest close to the fitness maxima.
Many such features as the stationary population density
in the fitness landscape or the mutation rate under which a population can still
be concentrated around a fitness maximum have been addressed
\cite{eigen:1977aa,eigen:1989aa,wilke:2005aa,nowak:1992aa}.
An important question is how a population 
evolves towards a fitness peak via several intermediate states.
If the intermediate states have the same fitness as the initial state, then
evolution to higher fitness states is neutral at first and thus poses no significant problems \cite{nimwegen:2000}.
If the intermediate states have lower fitness than the initial state, then a fitness valley has to be overcome and it is more difficult 
to reach the fitness peak \cite{weinreich:2006aa, poelwijk:2007aa}.
In this case, population stuck on a local peak cannot escape by natural selection alone, 
because there is no evolutionary trajectory with successively advantageous mutations.
Instead, neutral genetic drift becomes important.

Here, we consider the dynamics of these systems from a different perspective. 
We address the average time a population needs to transfer from one peak to another one. 
For small mutation rates and finite populations, we calculate this average time analytically. 
When mutation rates are high, we can describe the system by a set of differential equations and 
obtain the relevant times from a numerical integration of the differential equations. 
In this framework, the relevant question is how fast a population evolves \cite{traulsen:2007ee}.

In particular, we can address the question whether a population evolves faster from one
peak to another via $d$ mutations if
\begin{itemize}
\item[(i)] mutations have to occur in a certain order, i.e.\ only a single evolutionary trajectory is available
or
\item[(ii)] the order of the mutations does not matter, i.e.\ there are $d!$ evolutionary paths.
\end{itemize}

In the simplest case the intermediate fitness values are identical in both the cases and equal to that of the initial state.
Thus the only difference remaining is the number of available paths.
When the order of mutations is not fixed then multiple paths are available and the evolutionary dynamics will be faster when compared to a single path.
We can then ask the question: Does a population evolve faster on a narrow ridge or a broad valley?
This implies that we move away from the simplest case mentioned above and decrease the fitness in the intermediate states of the multiple paths compared to the fitness in the intermediate states of the single path.
We show how the pace of evolution depends on the depth of the valley, the number of intermediate states and the size of the population.

In general, evolutionary dynamics depends crucially on the size of the population. 
In a small population a single mutation will typically reach fixation or extinction before another mutation can arise. The population moves as a whole step by step on the fitness landscape.
For large populations, even for small mutation rates usually multiple types arise at the same time. 
This results in a non-zero population density in many states at the same time. 
For intermediate mutation rates, the population can either move stepwise across the fitness landscape or move several stpdf 
without getting concentrated in one of the intermediate states. This phenomenon has been termed stochastic  tunneling \cite{iwasa:2004aa}.
If the mutation rates are too small, tunneling does not occur because it is unlikely that a second mutation arises
before the first one has reached fixation or has gone extinct. 
If the mutation rates are high, tunneling occurs trivially, because the system can be approximated by differential equations for the densities in the different states. 
These different scenarios including the limiting cases of stepwise evolution (typical for small populations) and continuous evolution (typical for large populations) can also be observed when the population size is kept constant, but the mutation rates are increased.
For computer simulations increasing the mutation rate is more convenient than simulating huge populations for moderate mutation rates.

One important example for an evolutionary process in which the timescales are of crucial importance is the somatic evolution of cancer
\cite{frank:2007aa}.
Cancer progression has been investigated mathematically since the 1950s \cite{fisher:1959, nordling:1953, armitage:1954}.
Of special interest are the tumor suppressor genes \cite{knudson:1971aa,michor:2004aa}.
In a normal cell, there are two alleles of the tumor suppressor gene.
The mutation in the first allele is neutral if the second wild-type allele can sufficiently perform the function.
Inactivation of both the alleles confers a selective advantage to the cell and can lead to cancer progression.
This is an example in which the order of mutations does not matter. 
Many cancers also require certain particular mutations that initiate
cancer growth and pave the way for the accumulation of further mutations \cite{vogelstein:2004aa}.
Recently, it has been shown that after cancer initiation, a large number of different mutations may be
involved in cancer progression \cite{sjoeblom:2006sj,wood:2007ri,jones:2008ty,jones:2008ec}.
So far, it is unclear if the mutations have to occur in a specific order or if there is more
variation in the order \cite{beerenwinkel:2007aa,gerstung:2008aa}.

For simplicity, we consider only very simple fitness landscapes here in which the fitness in all the intermediate states is identical. 
In natural systems, these fitness values will differ and also the mutation rate may not be constant. 
In addition, sometimes the order of mutations will matter and sometimes, it will not.
Thus, sometimes a particular mutation will be a prerequisite to obtain a new function, but sometimes new mutations do not require any prerequisites. 
For example, this is the case in the evolution of resistance to $\beta$ lactam antibiotics studied by \cite{weinreich:2005aa} and \cite{weinreich:2006aa}.
However, here we focus on a very simple model to highlight the general aspects of the dynamics by analytical and numerical considerations.

This paper is organized as follows.
We begin with the description of the two ways to order the mutations, the single path and the hypercube. 
We then derive analytical approximations of the fixation times for small mutation rates and discuss the effect of the different parameters on the fixation times. 
Next, we address the dynamics for intermediate and high mutation rates.
Finally, we explore biological examples which can be modeled using this approach.

\section{Model}
To model evolutionary dynamics in a haploid population of size $N$, 
we use the Moran process \cite{nowak:2006bo,moran:1962ef}.
In each time step, one individual is selected at random, but proportional to fitness. 
It produces one offspring, which replaces a randomly chosen individual.
In one generation, each individual reproduces on average once. 
During reproduction, mutations occur with probability $\mu$.
We are interested in the time it takes until $d$ mutations reach fixation in the population,
starting from a homogeneous population in the initial state without any mutants.
Moreover, we aim to explore the dynamical features of this process. 
We restrict ourselves to two different cases that allow the derivation of some analytical results.

\subsection{Single path}
If the mutations can occur only in a particular order, we have a single evolutionary path, 
see Fig.\ 1 for an illustration.
Individuals in the initial state have fitness $r_0=1$ and individuals in the final state have fitness  $r_d>1$. 
It is instructive to characterize an individual by a string of $d$ sites, which can either be wild-type or mutated.
If the order of mutations is fixed, then a particular mutation requires another particular mutation as a prerequisite.
For simplicity, we assume that all the $d-1$ intermediate states have the same fitness $r_j=s<r_d$
($j=1, \ldots, d-1$).
For $s<1$, the joint effect of the set of mutations make up for the loss of fitness caused by the individually deleterious mutations.
This can be considered as a very special case of epistasis \cite{weinreich:2005ab}.

\subsection{Hypercube}
If the order of mutations does not matter, evolutionary dynamics takes place on a hypercube in $d$ dimensions 
cf.\ Fig.\ 1.
Thus, there are $2^d$ different types of individuals. 
In the initial state, we have $d$ possible mutations.
In the next step, $d-1$ mutations are available. 
Consequently, we have $d!$ possible paths to fixation.  
Again, we assume $r_0=1$ and $r_d>1$. Further,  all individuals with some, but not all mutations have fitness $s<r_d$.

\begin{figure}[h]
\begin{center}
\includegraphics[width=1.0\columnwidth]{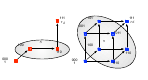}
\caption{
The order of mutations determines the 
geometry for evolutionary dynamics, shown here for $d=3$ sites (e.g. genes, nucleotide sites etc.).
If mutations can only occur in a particular order, 
only a single path is available (left). 
If the order of mutations is arbitrary, evolutionary 
dynamics occurs on a hypercube (right).  
The initial states have fitness $1$ and the final states fitness $r\geqslant1$. 
All intermediate states are assumed to have the same fitness $s<r$.
States are labeled by bit-strings, $0$ is an wild-type site and $1$ is a mutated site.
\label{fig:scenario}
}
\end{center}
\end{figure}

If the mutation probability is small, we do not need to make specific assumptions on the mutation process.
But when the mutation probability increases, we can no longer be certain that only a single mutation occurs during reproduction. 
For simplicity, we do not consider the possibility of backward mutations.
Although back mutations are often relevant, especially to escape from evolutionary dead ends \cite{dePristo:2007pi}, it is not straightforward to define the speed of evolution in a system with backward mutations. 
This is due to the fact that for sufficiently high mutation rates, fixation in the final state might never occur.
Other definitions of the end state of the system become arbitrary to a certain extend.
The probability $u_{m \to m+k}$ that the offspring of an individual with $m$ mutations has $m+k$ mutations ($m \leq m+k\leq d$) is 
\begin{align}
u_{m \to m+k} = \binom{d-m}{k} \mu^{k} (1-\mu)^{d-m-k}.
\label{mutations}
\end{align}
This equation is valid for the hypercube, where the order of mutations does not matter. 
Here, $\binom{d-m}{k}$ is the number of different types of mutants with $k$ additional mutations, $ \mu^{k}$ is the probability that mutations occur at $k$ sites and  $(1-\mu)^{d-m-k}$ is the probability that no mutation occurs at the remaining $d-m-k$ sites. 
For the single path, there is only one possibility to arrange the $m+k$ mutations. 
Thus, for $k>0$, $u_{m \to m+k}$ is identical to Eq.\ \eqref{mutations}, except that 
the binomial factor has to be dropped. 
The probability $u_{m \to m}$ that no mutation occurs follows from normalization,
$u_{m \to m} = 1-\sum_{k=1}^{d-m} u_{m \to m+k}$.
Our analytical calculations for small mutation rates as well as the considerations for high mutation rates are 
independent of the precise form of the mutation rates. 
However, we need to specify the form of the mutation probabilities to perform our numerical simulations for intermediate and high mutation rates. 

\section{Small mutation rates}

\subsection{The pace of evolution for small mutation rates }

For small mutation probabilities, double mutations can be neglected.
Since mutations occur rarely, 
we can calculate the average time until $d$ mutations are fixed in
the population analytically. Let us first address the evolutionary dynamics when mutants with fitness $r_m$ are already present in a resident population of fitness $r_w$, but no new mutations occur. 
This scenario is relevant when mutation rates are sufficiently small. 
The probability to increase the number of mutants from $j$ to $j+1$ is 
\begin{align}
T^+_j = \frac{r_m j}{ r_m j + r_w (N-j)}  \frac{N-j}{N}.
\end{align}
Similarly, the number of mutants decreases from $j$ to $j-1$ with probability 
\begin{align}
T^-_j = \frac{r_w (N-j)}{ r_m j + r_w (N-j)}  \frac{j}{N}.
\end{align}
The probability that $k$ mutants take over the entire population is given by
\cite{nowak:2006bo,karlin:1975xg,ewens:2004qe,crow:1970ck}
\begin{align}
\phi_k \left( \frac{r_m}{r_w} \right) 
=\frac{1+\sum_{i=1}^{k-1}{\prod_{j=1}^{i}{\frac{T^-_j}{T^+_j}}}}{1+\sum_{i=1}^{N-1}{\prod_{j=1}^{i}{\frac{T^-_j}{T^+_j}}}}
= \frac{1- \left(\frac{r_w}{r_m}\right)^{k}  }{1- \left(\frac{r_w}{r_m}\right)^{N}}.
\end{align}
If a mutant reaches fixation, the average number of generations 
this process takes is given by \cite{goel:1974aa,antal:2006aa}
\begin{align}
\tau_{\rm fix} \left( \frac{r_m}{r_w} \right) =\frac{1}{N}
\sum_{k=1}^{N-1}
\sum_{l=1}^{k}
\frac{\phi_{l}}{T_{l}^{+}}
\prod_{m=l+1}^{k} \frac{T^-_m}{T^+_m} .
\end{align}
For a neutral process with $r_m = r_w$, this reduces to $\tau_{\rm fix} = N-1$.
For sufficiently large $N$, this is the maximum conditional fixation time of a mutant. Even for disadvantageous mutants ($r_m < r_w$) the conditional fixation time is smaller than $N-1$ 
\cite{antal:2006aa}.
Since there are $\mu N$ mutations per generation, the time between two mutations is  
$\frac{1}{\mu N}$. Thus, for $\mu \ll N^{-2}$ a mutant reaches fixation before the next one arises
and mutations will not occur when a mutant is already present. Thus the population evolves by a process where the mutations occur one after the other, which has been termed periodic selection \cite{atwood:1951aa} and theoretically described as the strong-selection weak-mutation regime \cite{gillespie:1983aa,gillespie:2004bo}.

The total time $\tau$ until a mutation reaches fixation in a population is the sum of the waiting time until a successful mutant occurs and the fixation time of the mutant
$\tau = \tau_{\rm wait} +\tau_{\rm fix}$. 
The waiting time is the inverse of the mutation rate divided by the probability that a particular mutant is successful,
\begin{align}
\tau_{\rm wait} \left( \frac{r_m}{r_w} \right)  = \frac{1}{\mu N} \frac{1}{ \phi_1\left( \frac{r_m}{r_w} \right) }.
\end{align}
For $\mu \to 0$, we have  $\tau_{\rm wait}  \to \infty$, but $\tau_{\rm fix}$ remains approximately constant.
Thus, $\tau \approx  \tau_{\rm wait} $ for small mutation rates. 
In principle, we could calculate $\tau_{\rm fix}$ in the presence of mutations.
But since our approximation is only valid for small mutation rates, this will
be a minor correction.

For $\mu \ll N^{-2}$, the population is homogeneous most of the time. 
Only occasionally, a mutant arises and reaches fixation or goes to extinction. 
The total time until $d$ mutations are fixed in the population is the sum 
of the waiting times for the successful mutants plus the time
of the $d$ fixation events. 
For a single path with initial fitness $1$, intermediate fitness $s$ and 
final fitness $r$, we find for the total time $\tau^{S} $
\begin{align}
\tau^{S} &= \tau_{\rm wait} \left( s \right) + (d-2)  \tau_{\rm wait} \left( 1 \right) 
+   \tau_{\rm wait} \left( r/s \right) \\ \nonumber
&+ \tau_{\rm fix} \left( s \right) + (d-2) \tau_{\rm fix} \left( 1 \right)  + \tau_{\rm fix} \left( r/s \right). 
\end{align}
For small $\mu$, we have $ \tau_{fix} \ll \tau_{wait} $ and hence the total time can be approximated by 
\begin{align}
\tau^S &= \frac{1}{\mu}
\left[
\frac{1}{N \phi_1(s) } + d-2 + \frac{1}{N \phi_1(r/s) } 
\right] 
\label{sptime}
\end{align}
Consider now a ``fitness valley'', in which the intermediate states have fitness
$s<1$, but the final state has fitness $r>1$. To move from the fitness peak
in the initial state to the fitness state in the final state, first a disadvantageous mutation
has to be fixed in the population. Since $\phi_1(s<1) \ll \frac{1}{N}$, the waiting
time of such an event is very long. 
The waiting time for the neutral mutations, $ \tau_{\rm wait} \left( 1 \right) = \frac{1}{\mu}$
and the waiting time for a successful mutation, $ \tau_{\rm wait} \left( r/s \right) $
are significantly shorter. Thus, $\tau^S$ is dominated by $\frac{1}{\mu N \phi_1(s)}$
for $s<1$ and sufficiently high $N$ in a fitness valley. Fig.\ 2 shows a good agreement between exact numerical simulations and our analytical approximation for small mutation rates Eq.\ \eqref{sptime}.

If the order of mutations is arbitrary, evolutionary dynamics occurs
on a hypercube. In this case, the whole process will be faster,
as we have $d!$ possible paths instead of a single one. 
Now, the waiting times in the different states depend on the number
of mutations that are still available. 
For the total time $\tau^{H}$, we obtain,
\begin{align}
\tau^{H} &= \frac{1}{d} \tau_{\rm wait} \left( s \right) +  \sum_{k=1}^{d-2} \frac{1}{d-k}  \tau_{\rm wait} \left( 1 \right) 
+   \tau_{\rm wait} \left(  \frac{ r}{s} \right) \\ \nonumber
&+ \tau_{\rm fix} \left( s \right) + (d-2) \tau_{\rm fix} \left( 1 \right)  + \tau_{\rm fix} \left( \frac{ r}{s} \right) 
\end{align}
Note that the time of the fixations alone is identical for the hypercube and the
single path. 
Neglecting these fixation times for small $\mu$ ( as $ \tau_{fix} \ll \tau_{wait} $) yields
\begin{align}
\tau^H &= \frac{1}{\mu}
\left[
\frac{1}{d N \phi_1(s) } + \sum_{k=1}^{d-2} \frac{1}{d-k}  + \frac{1}{N \phi_1(\frac{ r}{s}) } 
\right].
\label{hctime}
\end{align}
Since 
$\frac{1}{d N \phi_1(s) }  < \frac{1}{N \phi_1(s) } $
and 
$\sum_{k=1}^{d-2} \frac{1}{d-k} < \sum_{k=1}^{d-2} 1 = d-2$ 
it is obvious that $\tau^H < \tau^S$, i.e.\ evolutionary dynamics
is faster if the order of mutations is arbitrary. 
For fitness valleys with $s<1$ and a large population size, $\tau^H$ is dominated by $\frac{1}{d \mu N \phi_1(s)}$.
As $d$ more mutations are available, this is always faster than the 
corresponding equation for a single path, see Fig.\ 2. 

\begin{figure}[h]
\begin{center}
\includegraphics[angle=-90,width=1.0\columnwidth]{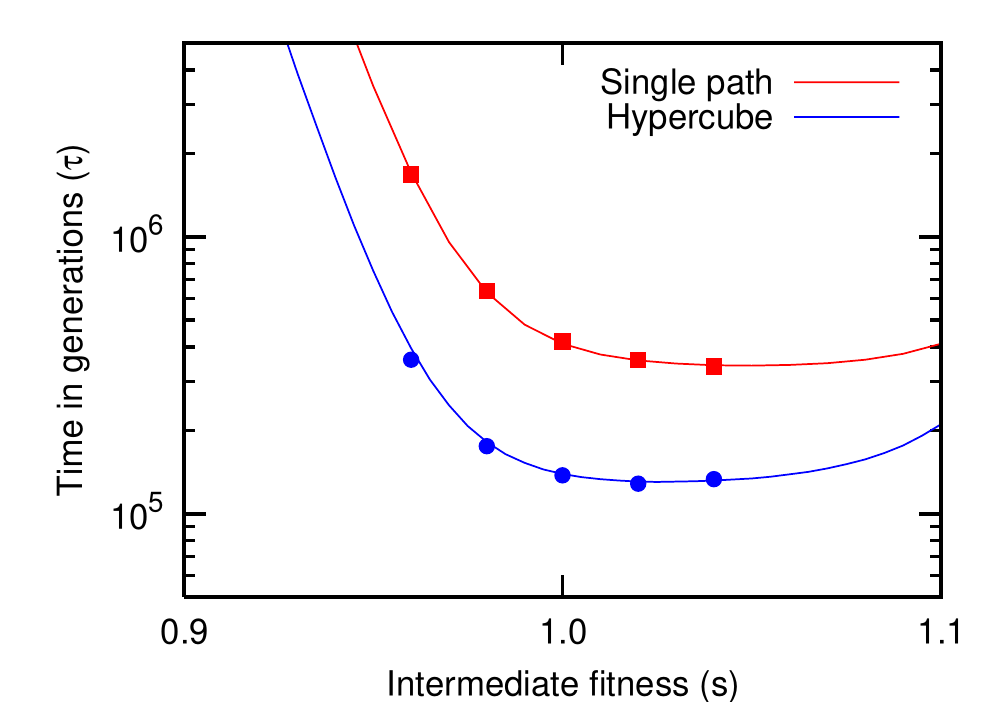}
\caption{
Fixation time for a single path (squares) and a hypercube (circles) with small mutation rates ($\mu \ll N^{-2}$) for
different intermediate fitness values. 
Evolutionary dynamics is always faster in the hypercube. 
Solid lines show the analytical approximation for small mutation rates, Eqs.\ \eqref{sptime}
and \eqref{hctime}. Numerical simulations shown by symbols agree well with the analytical approximation (population size $N=100$, mutation rate $\mu=10^{-5}$, $d=5$, $r=1.1$, simulations averaged over a 1000 realizations).
\label{fig:intermediate}
}
\end{center}
\end{figure}

\subsection{Thresholds of the waiting times }

Next, we derive expressions for some interesting thresholds of the waiting times in the limit of small mutation rates.
Since evolutionary dynamics is always faster if many paths are available, we
now compare a fitness valley in which many paths are available to a
single path in which the order of mutations is important, but fitness does not decrease in the course of evolution.
The basic question we address here is, whether it is faster to cross a broad valley or a narrow ridge in fitness space. 
In other words, we compare $\tau^{S}(s=1)$ to $\tau^{H}(s<1)$. 
Since we consider only small mutation rates $\mu$, we neglect the fixation times $\tau_{\rm fix}$ here, although they will not be identical in the two scenarios. 
For $s=1$, the single path is neutral. We decrease $s$ in the hypercube until we have
identical waiting times. This yields an implicit expression for $s$, 
\begin{align}
d-1+\frac{1}{N}\frac{1- \frac{1}{r^{N}}}{1- \frac{1}{r}}
=
\frac{1}{dN}\frac{1-\frac{1}{s^{N}}}{1-\frac{1}{s}}+\sum_{k=1}^{d-2}{\frac{1}{d-k}}+\frac{1}{N}\frac{1-(\frac{s}{r})^N}{1-\frac{s}{r}}
\label{numsol}
\end{align}
From this equation, we can numerically determine $s$ for any given $N$.
For large $N$, Eq. \eqref{numsol} simplifies to 
\begin{equation}
d (d-1)-\sum_{k=1}^{d-2}{\frac{d}{d-k}}
=
\frac{1}{N}\frac{1-\frac{1}{s^{N}}}{1-\frac{1}{s}}
\approx \frac{e^{N(1-s)-1}}{N(1-s)}, 
\label{reducednumsol}
\end{equation}
where we used $(1-x/N)^{-N} \to e^x$ for large $N$. 
Thus, the quantity $N(1-s)$ becomes constant for large $N$, see Fig.\ 3.
Thus, we can now say how broad and deep a fitness valley
has to be to lead to the same cumulative waiting time as a single neutral path.

\begin{figure}[h]
\begin{center}
\includegraphics[angle=-90,width=1.0\columnwidth]{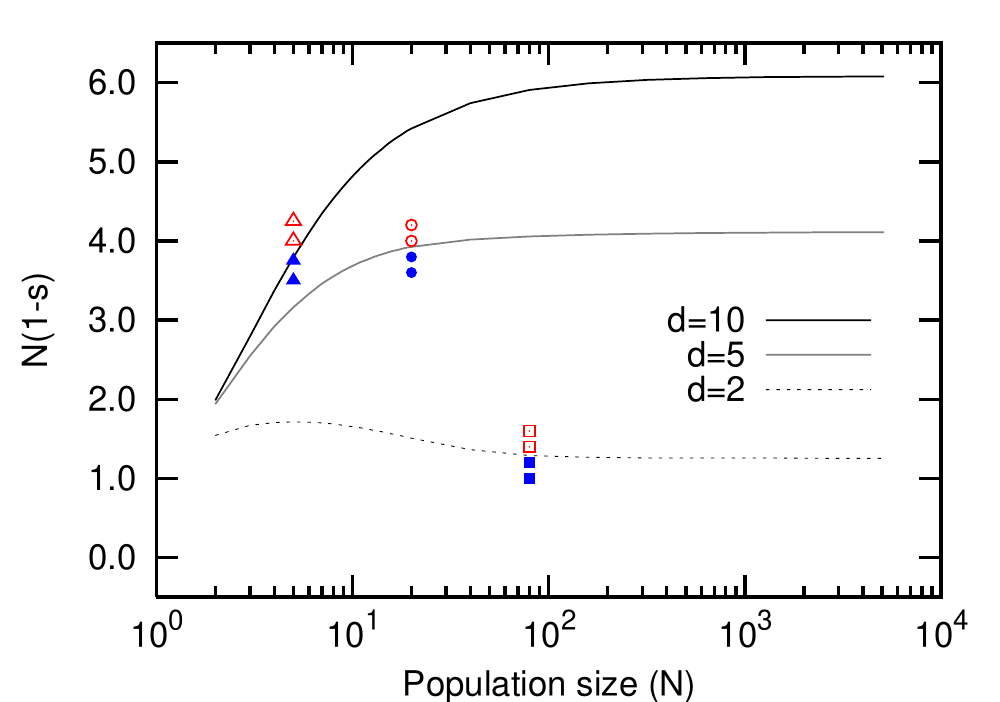}
\caption{
The figure shows the threshold values for which evolution on a hypercube with fitness $s<1$ in the intermediate states proceeds as fast as on a single neutral path with $s=1$. 
Full lines show $N(1-s)$ based on the numerical solution of Eq. \ref{numsol}. 
Above the lines, evolution proceeds faster on the hypercube. 
Below them, the neutral single path is faster. 
For $N \to \infty$, the lines converge to a constant, see Eq.\ \eqref{reducednumsol}. 
The symbols show the results from numerical simulations for $N=5$ and $d=10$ (triangles),
$N=20$ and $d=5$ (circles) as well as $N=80$ and $d=2$ (squares). 
Symbols are open (red) when the single path is faster and filled (blue) when the hypercube is faster
($\mu=10^{-5}$, $r=1.1$, simulations averaged over a 1000 realizations).
}
\label{fig:compare}
\end{center}
\end{figure}

Next, we address the effect of the intermediate fitness $s$, 
which has an important influence on the cumulative waiting time $\tau$. 
Fig.\ 2 shows how the waiting time decreases with increasing fitness in the intermediate states $s$.
If $s$ comes very close to the fitness in the final state, the waiting time increases. This increase is seen both in the single path and the hypercube. 
An increase in the intermediate state fitness will not always lead to a reduction in waiting times. Instead, the fixation times reach a minimum when the fitness growth is constant between any two consecutive states \cite{weinreich:2005tx,traulsen:2007ee}.
For the hypercube, the fastest trajectory will be steeper than on a single path: at first, many mutations are available and a big fitness increase is not necessary. Later, fewer mutations are available and thus, the fitness should increase faster. The precise form of the trajectory will in this case depend on the number of mutations $d$ and the population size $N$. 
We note that a similar reasoning can be applied to construct a fitness landscape that allows to cross a fitness valley fastest: the fastest trajectory has the same form regardless if a fitness peak is approached ($r>1$) or a fitness minimum is approached ($r<1$). Thus, the fastest way to cross a fitness valley is to descend to the minimum with exponentially decreasing fitness and to increase from the minimum again with exponentially increasing fitness.

Now, we turn to the effect of the intermediate fitness $s$ on the individual waiting times.
Eqs.\ \eqref{sptime} and \eqref{hctime} both consist of three terms each.
The first term denotes the time required to leave the initial state.
The second term is the time spent in moving through all the intermediate states.
This second term is independent of $s$, because the transitions are neutral.
The last term denotes the time required to reach the ultimate state from the penultimate state.
For small values of $s$, the probability to fixate the disadvantageous mutation is very small. 
Thus, the total time is dominated by the first term. When $s$ is increased to a threshold value $s_1$, 
then the time for leaving the first state is identical to the waiting time in the intermediate states. 
For the hypercube, $s_1$ is given by  $ \frac{1}{d} \tau_{\rm wait} \left( s_1 \right) =  \sum_{k=1}^{d-2} {\frac{1}{d-k}} \tau_{\rm wait} \left( 1 \right) $, which reduces to 
\begin{align}
\frac{1-\frac{1}{s_1^{N}}}{1- \frac{1}{s_1}} = d N \sum_{k=1}^{d-2}{\frac{1}{d-k}}.
\label{s1}
\end{align}
This equation can be solved numerically for specific values of $N$ and $d$.
For the single path, the right hand side of this equation has to be replaced by $N(d-2)$. 
For $s>s_1$, the time to cross the intermediate states is larger than the waiting time in the first state. 
On the hypercube, we can define a second threshold for which the waiting time in the first state is the same as the time required to reach the final state from the penultimate state. 
This arises because the effective mutation rate in state $0$ is $d$ times larger than the effective mutation rate in state $d-1$.
The threshold $s_2$ is given by 
$ \frac{1}{d} \tau_{\rm wait} \left( s_2 \right) =  \tau_{\rm wait} \left( \frac{r}{s_2} \right) $ or
\begin{align}
\frac{1}{d}\frac{1-\frac{1}{s_2^N}}{1-\frac{1}{s_2^1}}
=
\frac{1-\left(\frac{s_2}{r}\right)^{N}}{1-
\frac{s_2}{r}}
\label{s2}
\end{align}
Again, $s_2$ has to be determined numerically. For a single path, the factor $d^{-1}$ in 
Eq. \eqref{s2} has to be dropped. Thus, the threshold $s_2$ 
occurs for $s>1$ and is simply given by $s_2 = \sqrt{r}$.

The fixation time is also strongly influenced by the number of mutations $d$.
A larger $d$  increases the length of the path and usually also the fixation times. 
For the single path, this increase results only from the increase in the time required to cross the intermediate states, because the time for leaving the initial state and the time to reach the final state from the penultimate state are independent of $d$.
The time required to reach the ultimate state from the penultimate state is also independent of $d$ for the hypercube,  but the time required to leave the initial state decreases with increasing $d$. 
This is because as $d$ increases, there are more states available in the first error class and thus the effective rate of mutation out of the initial state increases.
As for the single path, the time to cross the intermediate states increases with $d$ in the hypercube.
For the hypercube, this interplay can lead to a non-monotonic dependence of the fixation time on $d$.
For example, for $N=100$ and $s=0.95$, the fixation time $\tau^H$ decreases with $d$ for $d <31$,
but it increases with $d$ for $d>31$. 
In contrast, the fixation time always increases monotonically with $d$ for the single path.

Increasing the fitness of the final state $r$  increases the advantage of the final state over the intermediate states. 
This will result in the decrease in the time required for the population to make the last move. 
Increasing $r$ has no effect on the time required to cross the intermediate states or the time required to move away from the initial state.
As a result, those two times remain constant even as $r$ increases, both in the single path and the hypercube.

\begin{figure}[t]
\begin{center}
\includegraphics[angle=-90,width=1.0\columnwidth]{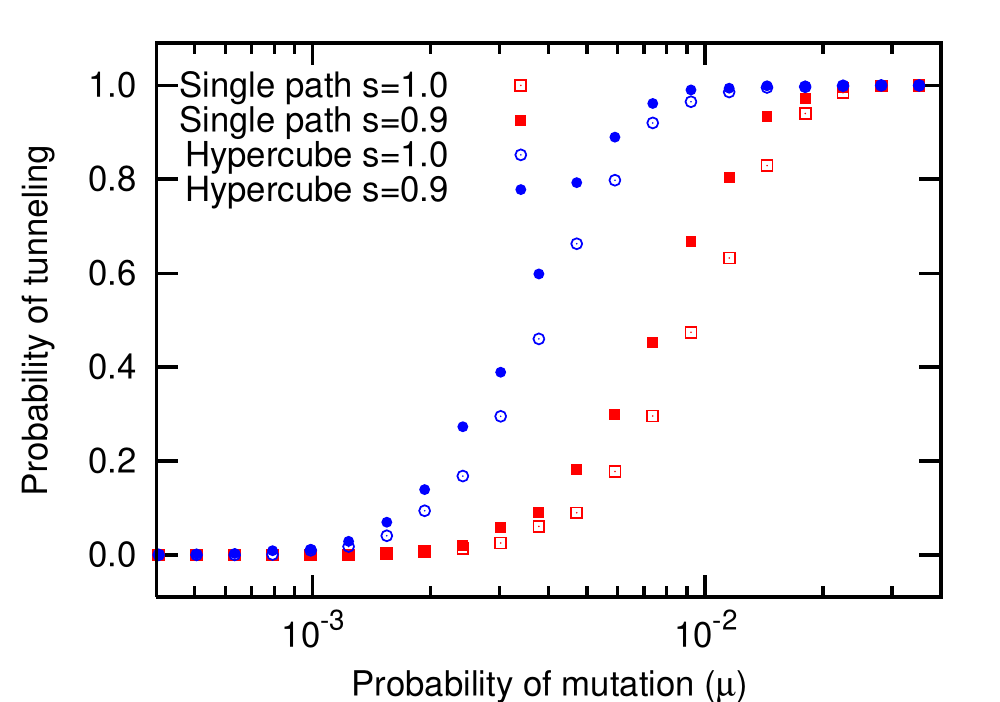}
\caption{
The probability of tunneling across the hypercube (circles, blue) is larger than in the single path (squares, red) due to the higher effective mutation rate. 
The tunneling across the valley denoted by the filled symbols ($s=0.9$) 
is always larger than  the probability of tunneling across a flat fitness landscape denoted by
open symbols ($s=1.0$). This arises from the fact that the stepwise accumulation of mutations
would involve the fixation of a disadvantageous mutation in the first step.
All symbols show the probabilities that the population tunnels at least across one state for the single path or at least one error class for the hypercube
($N = 100$, $d = 5$, $r = 1.1$, averaged over a 1000 realizations).
}
\label{valleytunnel}
\end{center}
\end{figure}

\section{Intermediate mutation rates}
The analytical approach is only valid as long as the mutation rate is small, $\mu \ll N^{-2}$.
For higher mutation rates, the population does not have to consist of at most two different types
at any time. Instead, $d$ mutations can be fixed in the population without sequentially fixing one after the other. This process has been termed stochastic tunneling and is of great importance in
the context of cancer initiation \cite{iwasa:2004aa,michor:2004aa, nowak:2004bb,beerenwinkel:2007aa}. 
Tunneling across fitness valleys is more likely than tunneling across a flat fitness landscape (see Fig.\ 4).
Even for $d=2$, the evolutionary dynamics is characterized by a doubly stochastic process, 
which makes analytical approaches tedious \cite{iwasa:2004aa}.
As discussed above, for $\mu N ^2\ll 1$ the population usually contains at most two different 
types. In this case, the probability of stochastic tunneling will be very small. 
On the other hand, for $\mu N > 1$, at least one mutant is produced per generation.
Thus,  the probability of stochastic tunneling approaches $1$. 
For $ N ^{-2} < \mu < N^{-1}$, the mutations are sometimes fixed sequentially and sometimes via stochastic tunneling. Fig.\ 5 shows how the tunneling probability increases from $0$ to $1$ in this interval.

\begin{figure}[b]
\begin{center}
\includegraphics[angle=-90,width=1.0\columnwidth]{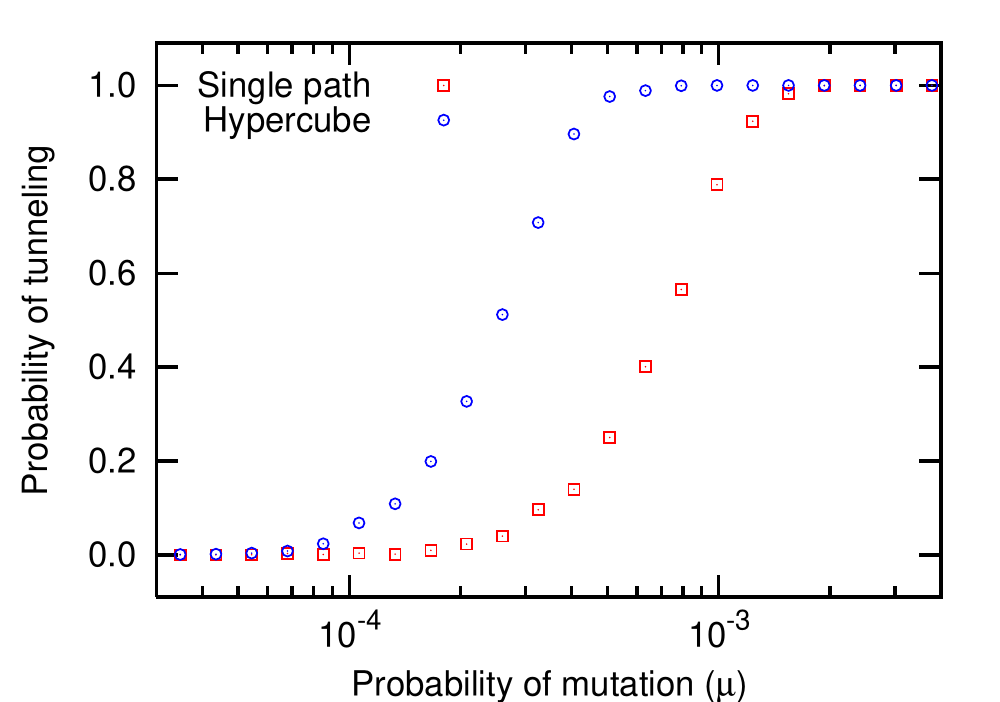}
\caption{
The probability of tunneling across a neutral hypercube (circles) is always higher than the probability of tunneling across a neutral single path (squares). Here, the probability that the system tunnels across at least one state or one error class is shown for $d=5$ and $N=1000$. 
As expected, for $N \mu>1$ the probability of tunneling approaches $1$. 
In contrast to our conservative estimate that tunneling can be neglected 
only as long as the mutation rate is below $N^{-2}$, even for mutation rates as large as $100 N^{-2}$ the probability of tunneling remains close to zero 
($s=1.0$, $r = 1.1$, averaged over a 1000 realizations).
}
\label{fig:tunnel}
\end{center}
\end{figure}

For intermediate mutation rates, it is likely that the population contains more than two different types. The types with beneficial mutations will compete for fixation. This process is termed clonal interference \cite{crow:1970ck,fisher:1930fi,muller:1932aa,gerrish:1998aa, park:2007aa}. Clonal interference has been considered to slow down adaptation, but recently it has been shown that it can have a positive influence on a rugged fitness landscape \cite{gerrish:1998aa,wilke:2004aa,jain:2007aa}.

The states in a single path can be characterized by the number of mutations.
In the hypercube, the states are characterized by the types of mutations that
have already occurred. Thus, there are many different types that have undergone
a specific number of mutations. 
However, all types that have already accumulated $k$ mutations can be pooled
into the error class $k$. 
The number of different types in the error class $k$ is given by 
$\binom{d}{k} = \frac{d!}{k! (d-k)!}$.

In a single path, a population is said to tunnel across a state
if it passes through a state without ever reaching fixation in that state.
Analogously, in a hypercube a population said to tunnel across an error class
if it passes through that error class without ever reaching fixation in it.
Within the error classes, tunneling can occur across individual states, but also across several states at once.
This means that the whole population passes only across that particular 
state and not across any other, without ever reaching fixation in that particular state.
Tunneling across an error class can also occur in a second way: the population can use all of the available states in the error class, 
but the total number of individuals in the error class never reaches $N$.
Thus, the probability of tunneling via the individual states is always lower than the probability of tunneling across the error classes. Fig.\ 6 shows the relation between the different probabilities of tunneling in the hypercube with respect to the rate of mutation $\mu$ for the special case $d=2$.

\begin{figure}[h]
\begin{center}
\includegraphics[angle=-90,width=1.0\columnwidth]{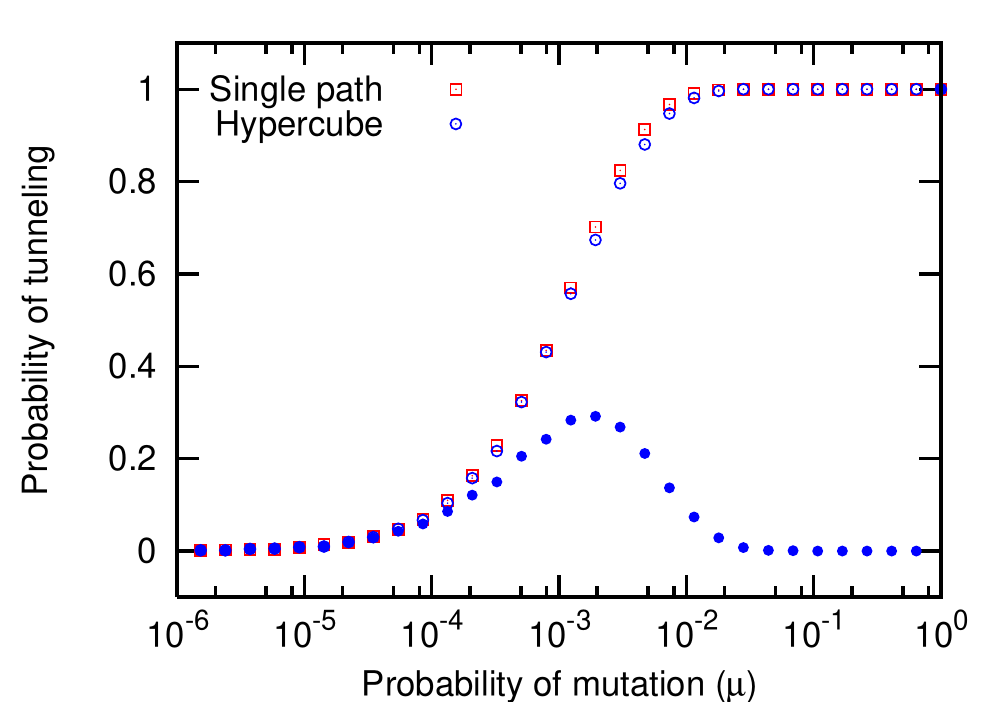}
\caption{
For $d=2$, the probability of tunneling across the intermediate state is 
slightly
higher in the single path (squares) than in the hypercube (open circles), shown for $N=100$ here. This is because the effective mutation rate into the intermediate state is twice as big in the hypercube, leading to a higher probability of fixation.
Filled
 circles show the probability to tunnel across individual states of the hypercube. For $\mu N>1$,
the system always tunnels. In the hypercube, both states are used for this. 
As expected, we need $\mu \ll N^{-2}$ for the tunneling probability to vanish 
($s=1.0$, $r=1.1$, averaged over a 1000 realizations).
}
\label{ectunnel}
\end{center}
\end{figure}

Due to higher effective rates of mutation, the probability of tunneling across a hypercube is expected to be greater than or equal to the probability of tunneling across a single path.
However, numerical simulations reveal that for $d = 2$ the probability of tunneling in a single path is higher than in the hypercube. This is a special case: 
for $d=2$ in a hypercube, the number of states into which the initial state can mutate into is $2$. The effective rate of mutations is thus twice as much as in the single path. The number of states which can be mutated into next is one, both in the single path and the hypercube. Thus the rate at which the individuals are pushed into the first state is higher in hypercube than in the single path while the rate of individuals being pushed out is the same. Thus there is a higher probability of reaching fixation in the first error class in a hypercube (see Fig.\ 6).
We only observe this effect for $d=2$, for $d>2$, the probability of tunneling  is higher in a hypercube than in a single path, as expected (see Fig. \ 4).

\section{High mutation rates}
For  $\mu N > 1$, the stochastic features of the dynamics become less important.
In this case, the system can be described by a set of $d+1$ deterministic differential equations for the fraction $x_k(t)$ of the population that has $k$ mutations \cite{jain:2007aa}.
Obviously, we have $\sum_{k=0}^{d} x_k(t) = 1$. 
Transitions out of state $0$ occur with probability 
$T_{0 \to} = (1- \frac{x_0}{\phi} u_{0 \to 0}  ) x_0$, 
where $\phi =  x_0 + (1-x_0-x_d) \, s+ x_d \, r$ is the average fitness of
the population. 
This includes all the reproductive events except for the one where a type $0$ is produced.
Transitions into state $0$ occur with probability 
$T_{\to 0} = \frac{x_0}{\phi} u_{0 \to 0}   (1-x_0)$.  
Thus, the fraction of individuals in the initial state follows the differential equation
\begin{align}
\dot x_0(t)
= \frac{1}{N}\left[
\frac{x_0}{\phi} u_{0 \to 0}   (1-x_0)
-
(1- \frac{x_0}{\phi} u_{0 \to 0}  ) x_0 
\right].
\label{0eq}
\end{align}
The probability that an offspring is of type $k$ is given by 
$\lambda_k = \sum_{j=0}^k \frac{x_j r_j}{\phi} u_{j \to k}$.
The difference between the hypercube and the single path 
only occurs in the quantity $u_{j \to k}$, which is given above for 
both cases. 
The sum in $\lambda_k$ is over all individuals with $k$ or less mutations and $r_j$ is the fitness of individuals with $j$ mutations. This leads to the differential equation for the fraction of individuals with $k$ mutations, 
\begin{align}
\dot x_k(t)
= \frac{1}{N}\left[ \lambda_k (1-x_k)- ( 1-\lambda_k) x_k \right],
\label{keq}
\end{align}
where $k=0, \ldots , d$. 
Of course, the special case $k=0$ recovers Eq.\ \eqref{0eq}.
This set of $d+1$ differential equations describes how the system moves from state $k=0$ to
state $k=d$. 
In general, only a numerical solution of this system of equations is feasible. 
While this allows us to infer details of the dynamics, our main interest is the time required for fixation of $d$ number of mutations. 
Thus, we solve the differential equation numerically using a standard Runge-Kutta algorithm \cite{press:2007aa}.  To find an equivalent to the fixation time in a stochastic simulation, we average between fixation ($x_d=1$) and the time when there are on average less than $1$ individuals outside the final state ($x_d=1-\frac{1}{N}$). 
Thus, the fixation time is the time when the solution of the differential equation crosses 
$x_d=1-\frac{1}{2N}$.

\begin{figure}[h]
\begin{center}
\includegraphics[angle=-90,width=1.0\columnwidth]{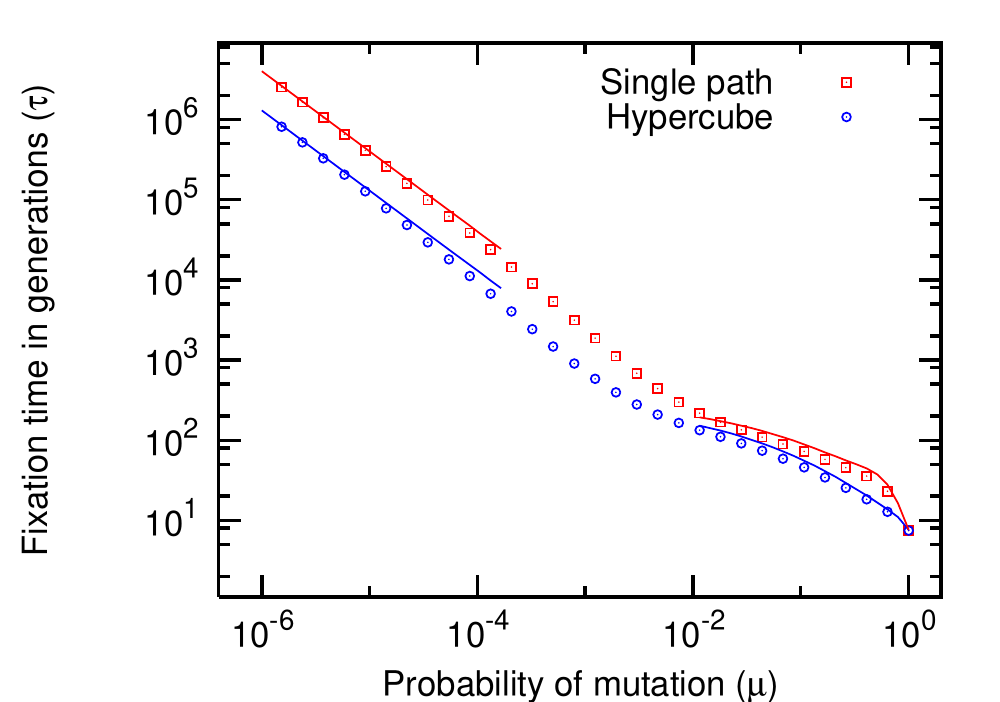}
\caption{
The fixation times decrease with increasing mutation rate. 
Fixation always occurs faster on the hypercube (circles) than in the single path (squares). 
For small mutation rates, mutations fixate sequentially and the fixation time can be well approximated by Eqs.\ \eqref{sptime} and \eqref{hctime}.
Here, the fixation times decrease as $\mu^{-1}$.
For high mutation rates, the system can be approximated by a set of deterministic differential equations and the simulation results for the fixation times
can be approximated based on the numerical solution 
of Eq. (\ref{keq}). 
In this case, fixation times decrease in general slower than $\mu^{-1}$ with increasing mutation rate 
(population size $N=1000$, $d=5$, $s=1$, 
$r=1.1$, averages over 1000 realizations).
}
\label{timeoverview}
\end{center}
\end{figure}

Fig.\ 7 shows an overview of the fixation times, covering the full range of mutation rates. For small mutation rates, we have sequential fixation of mutations and the time can be well approximated by Eqs.\ \eqref{sptime}
and \eqref{hctime}.
For high mutation rates, the numerical solution of Eq. (\ref{keq}) leads to a good approximation for the fixation times.

\section{Discussion}
\label{discussion}

We have determined the average time during which a population moves from a certain initial state to a final state of higher fitness.
The initial and the final states are separated by a fixed number of mutations $d$.
The mutations jointly confer a fitness advantage to the final mutant which can be represented by a peak in the fitness landscape.
If the intermediate mutations need to occur in a specific order for the evolution of the final mutant then it corresponds to the single path.
Otherwise, evolution occurs on a hypercube and there are $d!$ ways of reaching the final state.

We have explored the simplest system in which the fitness in all intermediate states is the same. 
As expected, the fixation times on a hypercube are shorter than on a single path, due to the presence of multiple paths available in a hypercube. 
This observation leads to the question: for which parameters does the hypercube show shorter fixation times than the single path, even with an added disadvantage?
The fitness in the intermediate states was then set to lower values than the ones in the single path. 
Up to a certain threshold value of the fitness of the intermediate states, the hypercube shows shorter fixation times than in the single path. The value of the threshold depends on the population size, 
total number of required mutations
and the fitness in the final state.

The fixation times for large populations largely depend on the fitness function and are qualitatively
independent of the order of mutations.
Let us first focus on a flat landscape: 
when the intermediate states have a fitness equal to the fitness of the initial wild-type, then for small mutation rates large populations have shorter fixation times than small populations. 
This is because the neutral rate of evolution does not depend on the population size.  But the waiting time for fixation of the last mutation becomes shorter with larger population size. 
For intermediate mutation rates, tunneling starts earlier in larger populations.
This leads to a marked decrease in the fixation time with larger population size. 
For high mutation rates, the time to fixation is no longer dominated by the time for the first mutant to reach the final state, but by the time until all individuals are in that state. Due to this, for high mutation rates the time required for fixation can be shorter in smaller population as compared to larger populations.
Next, we focus on fitness valley: If the fitness landscape consists of a valley with reduced fitness of the intermediate states, small populations have an advantage for small mutation rates, as they can easily leave the initial state and enter the valley. 
But for high mutation rates, large populations reach fixation faster, because they can explore states within and beyond the fitness valley more easily.

Our numerical simulations reveal that tunneling can be neglected even when the mutation rate
exceeds $N^{-2}$, at least by one order of magnitude.
Thus, Eqs.\ \eqref{sptime} and \eqref{hctime} provide good estimates for the
fixation times even in relatively large populations. 
Concrete values for fixation times are collected in Table\ 1. 
They reveal that even in long-term studies of experimental evolution, 
it is difficult to observe the consecutive fixation of neutral mutants \cite{cooper:2003aa}.
Consecutive fixation of advantageous mutants, however, is significantly faster.
For example, while Table\ 1 reveals a fixation time of $\sim 10^{11}$ generations on a single path
for $d=10$, $s=1$ and $N=10^6$, an optimal choice of the intermediate fitness values \cite{traulsen:2007ee} would lead to a fixation time of $\sim 10^7$ generations.

\begin{table}[h]
\begin{center}
\begin{tabular}{ccccc}
\hline
$N$	& \multicolumn{2}{c}{$d=3$}&\multicolumn{2}{c}{$d=10$}     \\
	& Single Path 	& Hypercube 	& Single Path & Hypercube \\
\hline
$10^{2}$		& $2.10999$	& $0.943325$	& $9.10999$	& $2.03896$\\
$10^{4}$	& $2.0011$	& $0.834433$	& $9.0011$	& $1.93007$\\
$10^{6}$	& $2.00001$& $0.833344$& $9.00001$	& $1.92898$ \\
\hline
\end{tabular}
\caption{
The time required for fixation of $d$ mutations in units of $10^{10}$ generations for a mutation rate of 
$\mu = 10^{-10}$
based on Eqs.\ \eqref{sptime} and \eqref{hctime}. The intermediate mutations are neutral, $s=1$.
For small mutation rates, the fixation times scale linearly with $\mu^{-1}$. 
For $N \to \infty$, the fixation time on the single path approaches $\mu^{-1}(d-1)$
and the fixation time on the hypercube approaches $\mu^{-1} \sum_{k=0}^{d-2} (d-k)^{-1}$. However,
the mutation rates have to decrease with increasing $N$ to make the approximation for the fixation times valid.    
 (initial fitness $1.0$ and final fitness $r=1.1$).}
\label{tabtimes1}
\end{center}
\end{table}

While we have focused on the simplest possible system which allows analytical approximations, 
experimental studies reveal of course a much higher complexity. 
\cite{weinreich:2006aa} studied experimentally the point mutations in the $\beta$-lactamase gene 
of bacteria. 
$\beta$ lactam antibiotics are commonly used, but the bacteria can develop resistance to the drugs.
Five point mutations in a particular allele of the $\beta$-lactamase gene increases the resistance of the bacteria to cefotaxime by a factor of $\sim100,000$.
Theoretically the mutations leading from the wild-type allele to the resistant allele can occur in $5! = 120$ ways. These can be represented by a hypercube of $d = 5$. 
But in only $18$ of the $120$ trajectories, the intermediate mutations are either neutral with respect to the initial state or beneficial.
Weinreich and colleagues have shown that these have the highest probability of realization. 
For all beneficial intermediates the fastest way to reach the final state would be when the relative fitness increase between any two consecutive mutations is constant \cite{traulsen:2007ee}, but usually 
in nature several different mutations are available and 
the population first evolves to states that provide the highest selective advantage.

In another experimental study the sequence space of the 5s rRNA of a marine bacterium, {\it{Vibrio proteolyticus}} was explored \cite{lee:1997}.
The sequences from {\it{Vibrio proteolyticus}} and {\it{Vibrio alginolyticus}} differ in only four positions.
All the possible intermediates were constructed by the authors and the fitness of each was calculated \cite{chao:1985aa}.
Two of the valid intermediates have a fitness lower than the initial wild-type. 
We have shown how such fitness valleys can be crossed by exploring the phenomenon of tunneling or multiple mutations (for high mutation rates). Thus, the population does not need to move in a Wrightian fashion (the whole population moving as a whole across the valley).

The theory discussed herein deals with basic evolutionary concepts which are important to the kind of biological examples described above.
More complex properties of the experimental studies like 
more general cases of
epistasis and compensatory mutations can easily be incorporated, but there is a huge number of possibilities. Even if we are only
interested in the ordering of fitness values, we can have up to $2^d!$ distinct epistatic patterns. 
Thus, one should rather focus on concrete systems instead.
For example, one could simulate the dynamics in a system with experimentally derived fitness values and mutation rates.
Not all the paths of a hypercube might be accessible for selection, but still some of them might prove to be significant depending upon the particular values of the parameters, such as fitness values and population size. 
Our goal here was to characterize the simplest features of the dynamics of a population crossing a fitness valley.
This approach can be helpful when more realistic scenarios are addressed. 

\section{Acknowledgment}
We thank Andrew Murray for initiating this project. C.S.G. and A.T. are grateful for financial support from the Emmy-Noether program of the DFG.
M.A.N. is supported by the John Templeton Foundation, the NSF/NIH joint program in mathematical biology (NIH Grant R01GM078986), the Bill and Melinda Gates Foundation (Grand Challenges Grant 37874), and J. Epstein.

\end{document}